\documentclass[preprint,showpacs,amsmath,amssymb,aps,prb]{revtex4}

\usepackage{graphicx}	
\usepackage{dcolumn}	
\usepackage{bm}			
\usepackage{here}

\begin{document}

\preprint{APS/123-QED}

\title{Spin-singlet superconductivity in the doped topological crystalline insulator Sn$_{0.96}$In$_{0.04}$Te}

\author{Satoki Maeda$^1$}
\author{Ryohei Hirose$^1$}
\author{Kazuaki Matano$^1$}
\author{Mario Novak$^2$}
\author{Yoichi Ando$^{3}$}
\author{Guo-qing Zheng$^{1,4}$}
\affiliation{$^1$Department of Physics, Okayama University, Okayama 700-8530, Japan}
\affiliation{$^2$Institute of Scientific and Industrial Research, Osaka University, Osaka 567-0047, Japan}
\affiliation{$^3$Physics Institute II, University of Cologne, K\"oln 50937, Germany}
\affiliation{$^4$Institute of Physics,  Chinese Academy of Sciences, and Beijing National Laboratory for Condensed Matter Physics, Beijing 100190, China}

\begin{abstract}
The In-doped topological crystalline insulator Sn$_{1-x}$In$_x$Te is a candidate for a topological superconductor, where a pseudo-spin-triplet state has been proposed.
To clarify the spin symmetry of Sn$_{1-x}$In$_x$Te, we perform $^{125}$Te-nuclear magnetic resonance (NMR) measurements in polycrystalline samples with 0$\leq x \leq$0.15.
The penetration depth calculated from the NMR line width  is $T$ independent below half the superconducting transition temperature ($T_{\rm c}$) in polycrystalline Sn$_{0.96}$In$_{0.04}$Te, which indicates a fully opened superconducting gap.
In this sample, the spin susceptibility measured by the spin  Knight shift ($K_{\rm s}$) at an external magnetic field of $\mu_0H_0$ = 0.0872 T decreases  below  $T_{\rm c}$, and $K_{\rm s}(T=0)/K_{\rm s}(T=T_{\rm c})$  reaches $0.36 \pm  0.10$, which is far below the limiting value 2/3 expected for a spin-triplet state for a cubic crystal structure.
Our result indicates that polycrystalline Sn$_{0.96}$In$_{0.04}$Te is a spin-singlet superconductor. 

\end{abstract}


\maketitle

\section{Introduction}
Topological insulators (TIs)  and  topological crystalline insulators (TCIs) are materials in which the bulk is insulating but the surface hosts metallic states due to non zero topological invariants of the bulk band structure\cite{Hasan,Qi,Hsieh,Tanaka,Ando}.
A TI requires time-reversal symmetry, while a TCI requires certain symmetries in crystal structure such as mirror symmetry.
Recently, superconductivity realized in carrier doped TIs or TCIs has attracted great interests, as it can be topological. 
A topological superconductor is analogous to TI or TCI in that the superconducting gap function has a nontrivial topological invariant\cite{Qi,Fu,sato}.
Vast efforts have been devoted to establishing topological superconductivity  with time-reversal symmetry in a bulk material,  but the progress had been slow 
until the recent discovery of a pseudo-spin-triplet, odd-parity superconducting state \cite{Matano} in the
 doped TI, Cu$_x$Bi$_2$Se$_3$ \cite{Hor}. 

SnTe with NaCl-type crystal structure is a TCI \cite{Hsieh,Tanaka} and shows superconductivity upon Sn-vacancies or In doping \cite{Hulm,Allen,Erickson}.
A quasi-localized impurity bound state due to In doping was recently evidenced by $^{125}$Te-NMR measurements \cite{MaedaSnTe},
which forms the background electronic state responsible for superconductivity \cite{Shelankov,Haldo}.
Point-contact spectroscopy performed on clean single crystals of Sn$_{0.955}$In$_{0.045}$Te found a zero-bias conductance peak, 
which was taken as a signature of unconventional superconductivity\cite{Sasaki}.
Specific heat \cite{Mario}, thermal conductivity \cite{He}, and $\mu$SR \cite{Saghir} have revealed a fully-opened  superconductiviting gap.
Combining these results, a fully-gapped pseudo-spin-triplet state was theoretically proposed \cite{Hashimoto}.
However, since the spin symmetry of Cooper pairs is unexamined, Knight shift measurements by nuclear magnetic resonance (NMR)
that can probe  the spin susceptibility below $T_{\rm c}$ are highly desired.

In metals, the Knight shift ($K$) contains two contributions as $K = K_{\rm chem} + K_{\rm s}$, where $K_{\rm chem}$ is the chemical shift, which is composed of  contributions due to orbital susceptibility and diamagnetic susceptibility of closed inner shells,  and $K_{\rm s}$ is due to  spin susceptibility.
The temperature variation of $K_{\rm s}$ below $T_{\rm c}$ depends on the spin symmetry of the Cooper pairs.
For a spin-singlet superconductor with a weak spin-orbit interaction, $K_{\rm s}$ decreases below  $T_{\rm c}$ and vanishes at $T$ = 0 K.
On the other hand, the $K_{\rm s}$ of a spin-triplet superconductor depends on the detail of the $d$ vector that describes the  paired spins.
The $d$ vector is perpendicular to the plane in which the parallel spins lie, and when this vector is pinned to a special
direction of the lattice, the $K_s$ is invariant across $T_c$ for a magnetic field applied perpendicular to the $d$ vector, while it
decreases for a magnetic field parallel to the $d$ vector.
This was indeed observed for the first time in  Cu$_x$Bi$_2$Se$_3$ \cite{Matano}.
%
For the fully gapped spin-triplet state proposed for Sn$_{1-x}$In$_x$Te \cite{Hashimoto}, $K_{\rm s}$ will decrease in a certain direction if the spins are well fixed to the lattice, as in Cu$_x$Bi$_2$Se$_3$ \cite{Matano}.
In the case of polycrystalline samples with a cubic structure, where $K_{\rm s}$ is an average over all directions, at most one-third of the $K_{\rm s}$ can be reduced at $T$=0. 
Therefore, measurement of the temperature variation of  $K_{\rm s}$ allows one  to determine the spin pairing symmetry. 

In this paper, we report $^{125}$Te-NMR measurements of polycrystalline Sn$_{1-x}$In$_x$Te.
First, we determine the quantity $K_{\rm chem}$ 
 using the relationship between $K$ and the spin-lattice relaxation time ($T_1$) of Sn$_{1-x}$In$_x$Te with various $x$'s.
 Then we measured   the $K_{\rm s}$ for Sn$_{0.96}$In$_{0.04}$Te down to $T$=0.1 K under the very small magnetic field of $\mu_0H_0$ = 0.0872 T.
 The obtained result indicates a spin-singlet pairing.

\section{EXPERIMENTAL PROCEDURE}
Polycrystalline samples of $x$ = 0, 0.05, 0.1, and 0.15 were synthesized by a sintering method at Okayama as described in the previous paper \cite{MaedaSnTe}.
An effectively polycrystalline sample of $x$ = 0.04 was synthesized by a melt-growth technique at Osaka.
This sample was initially attempted to be grown   as a big single crystal, but Laue diffraction showed that it consists of  many crystallites.
The $T_{\rm c}$ was determined by measuring the inductance of the NMR coil. 
NMR measurements were carried out by using a phase-coherent spectrometer.
NMR spectra under an external magnetic field $\mu_0H_0$ = 5 T were obtained by integrating the spin echo intensity by changing the resonance frequency ($f$).
In order to minimize the reduction of $T_{\rm c}$ by the applied field, most of the measurements for $x$=0.04 were performed at the small field of $\mu_0H_0 $= 0.0872 T,
under which the NMR spectra   were obtained by a fast Fourier transform of the spin echo.
The $T_1$ was measured by using a single saturating pulse, and determined by fitting the recovery curve of the nuclear magnetization to a single exponential function, $(M_0-M(t))/M_0 = \exp(-t/T_1)$, where $M_0$ and $M(t)$ are the nuclear magnetization in the thermal equilibrium and at a time $t$ after the saturating pulse.
Measurements below 1.4 K were carried out with a $^3$He-$^4$He dilution refrigerator.
After completion of  all the NMR measurements, the large sample of $x$=0.04 was crushed into several pieces and Hall coefficient measurements were performed on them. The Hall coefficient shows a  distribution of 30\% from piece to piece, but the averaged value indicates that the averaged $x$ over the sample is 0.04.  

\section{Results}

\begin{figure}[H]
	\begin{center}
		\includegraphics[width=8cm]{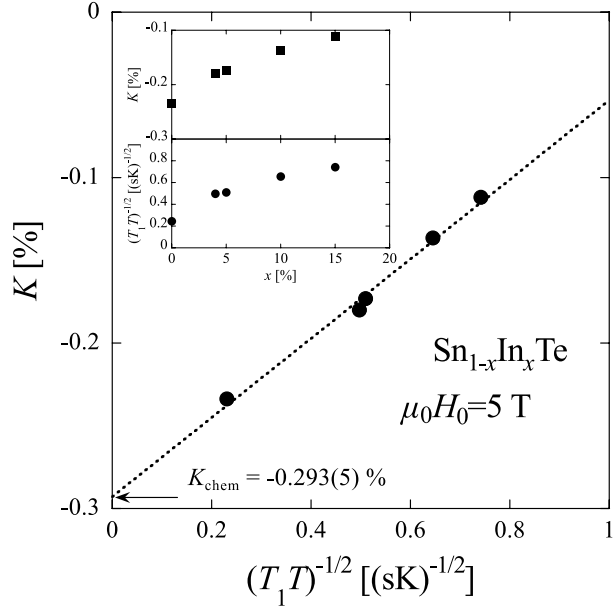}
		\caption{$K - (T_1T)^{-1/2}$ plot for various $x$'s under $\mu_0H_0$=5 T. Inset: The $x$ dependence of the $K$ and $(T_1T)^{-1/2}$ measured at the peak position of the spectrum. Smooth evolution of the two physical quantities indicates that the real doping level changes smoothly with $x$.}
		\label{Korb}
	\end{center}
\end{figure}

We first explain how we obtained $K_{\rm chem}$. In a normal metal, both $K_{\rm s}$ and the quantity $(T_1T)^{-1/2}$ are proportional to the density of states at the Fermi level [$N(E_{\rm F})$] and  $K_{\rm s}$ and  $T_1$ satisfy the Korringa relation $T_1TK^2_{\rm s}=\frac{\hbar}{4\pi k_B}(\frac{\gamma_e}{\gamma_n})^2$, where $\gamma_{e(n)}$ is the gyromagnetic ratio of the electron (nucleus).  This was recently found to be true in this system under a relatively high field ($\mu_0H_0$ = 5 T) \cite{MaedaSnTe}.
The inset in Fig. \ref{Korb} shows the $x$-dependence of the $K$ and the $(T_1T)^{-1/2}$ measured at the peak position of the spectrum.
The $K$ and the $(T_1T)^{-1/2}$ increased with an increase in $x$, which means an increase in $N(E_{\rm F})$ with increasing $x$.
As shown in Fig. \ref{Korb}, $K$ and $(T_1T)^{-1/2}$ show a good linear relationship with $x$ as an implicit parameter. Thus the $K_{\rm chem}$ can be determined as an intercept in a $K$ - $(T_1T)^{-1/2}$ plot.
As shown in Fig. \ref{Korb}, by extrapolating the data to the origin where $(1/T_1T)^{-1/2} = 0$, $K_{\rm chem} = -0.293 \pm 0.005$\% is obtained. The negative value of $K_{\rm chem}$ is due to the large  diamagnetism of the inner shells.

\begin{figure}[H]
\begin{center}
\includegraphics[width=8cm]{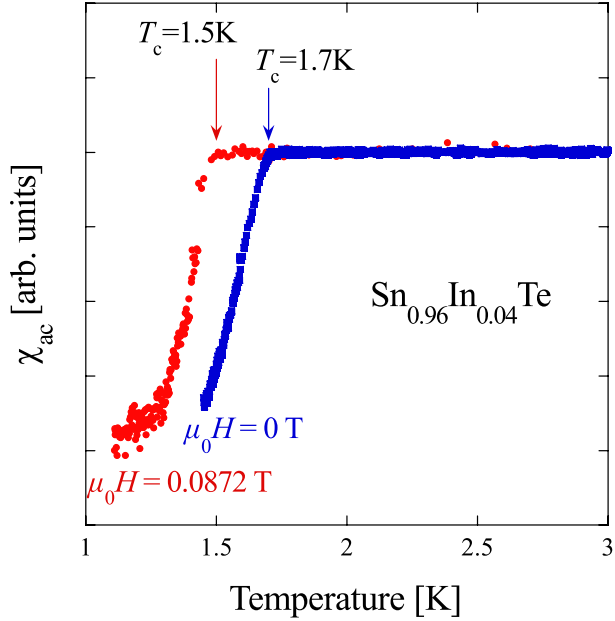}
\caption{(Color online) Temperature dependence of the $\chi_{\rm ac}$ for Sn$_{0.96}$In$_{0.04}$Te at $\mu_0H$=0 and 0.0872 T. 
}
\label{chi}
\end{center}
\end{figure}

Next, we discuss the  result in the superconducting state.
Figure \ref{chi} shows the temperature dependence of the ac susceptibility ($\chi_{\rm ac}$) for Sn$_{0.96}$In$_{0.04}$Te, which showed superconductivity at 1.7 K under $\mu_0H_0 = 0$ T and at 1.5 K under $\mu_0H_0 = 0.0872$ T.
The $T_{\rm c}$ at $H_0 = 0$ T was significantly higher than the reported value for $x$ $\sim$ 0.04 \cite{Sasaki,Gu}, which is commented on later. 
It is reported that the upper critical field $H_{\rm c2}$ for  Sn$_{1-x}$In$_x$Te with a high In contents is well fitted by  the parabolic formula \cite{Saghir}, $H_{\rm c2}(T) = H_{\rm c2}(0)[1-(T/T_{\rm  c})^2]$.
Using this relation,  $\mu_0H_{\rm c2}$ = 0.43 T is obtained. 
On the other hand, by using the Werthamer-Helfand-Hohenberg theory \cite{WHH},   $\mu_0H_{\rm c2}$ = 0.53 T is obtained from the initial slope of $H$ vs $T_{\rm c}$.

\begin{figure}[H]
\begin{center}
\includegraphics[width=8cm]{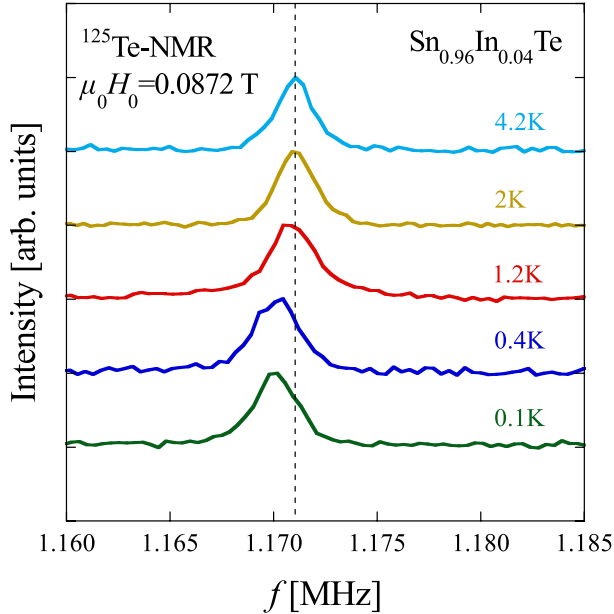}
\caption{(Color online) $^{125}$Te-NMR spectra for Sn$_{0.96}$In$_{0.04}$Te at various temperatures under $\mu_0H_0$ =0.0872T.}
\label{Spectra}
\end{center}
\end{figure}

\begin{figure}[H]
\begin{center}
\includegraphics[width=8cm]{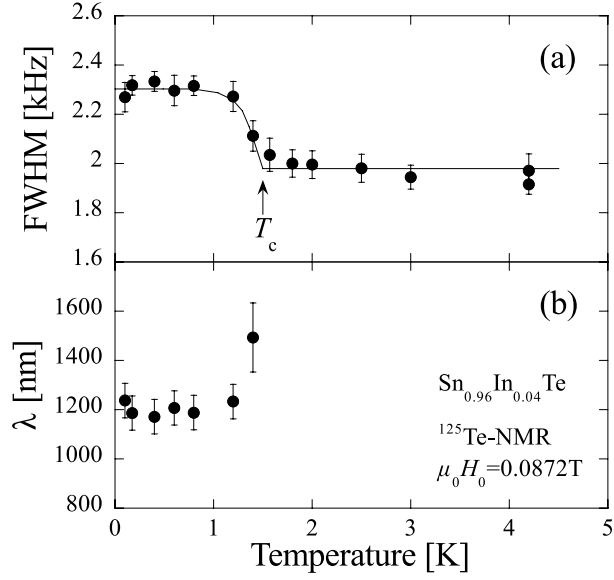}
\caption{(a)Temperature dependence of the FWHM for Sn$_{0.96}$In$_{0.04}$Te under $\mu_0H_0$ = 0.0872 T. The curve is a guide for eye. (b)Temperature dependence of the penetration depth $\lambda$ calculated from the FWHM.}
\label{FWHM}
\end{center}
\end{figure}

Figure \ref{Spectra} shows the temperature dependence of the $^{125}$Te-NMR spectrum for Sn$_{0.96}$In$_{0.04}$Te under $\mu_0H_0$ =0.0872 T.
The peak is temperature-independent above $T_{\rm c}(H)$ = 1.5 K, but shifts to a lower frequency  with decreasing temperature.
Figure \ref{FWHM}(a) shows the temperature dependence of the full width at half-maximum (FWHM). 
The FWHM increases below $T_{\rm c}$, due to  a magnetic-field distribution in the vortex state.
It is related to the penetration depth ($\lambda$)
as\cite{Brandt}
\begin{equation}
\sqrt{ {\rm FWHM}^2(T)-{\rm FWHM}^2(T_{\rm c})} = 0.0609\gamma_n\frac{\phi_0}{\lambda^2 (T)}.
\end{equation}

The $\lambda(T = 0) \sim 1,200$ nm was obtained from the above equation, which is larger than the $\lambda = 542$ nm reported by muon-spin spectroscopy for a sample with a higher In concentration ($x$ = 0.4, $T_{\rm c}$=4.69K)\cite{Saghir}.
Since the $\lambda$ is proportional to the carrier concentration $n$  as $-1/2$ ($\lambda \propto n^{-1/2}$) (Ref. \cite{deGennes}), the difference in $\lambda$ between $x$ = 0.04 and 0.4 is most likely due to the difference in carrier concentration.

As shown in Fig. \ref{FWHM}(b), $\lambda$ is $T$-independent below 0.5 $T_{\rm c}$, which indicates that the superconducting gap is  fully opened.
In a superconductor with nodes, $\lambda$ is proportional to $T^n$ ($n \geq 1$) at low temperatures.
Our result  is consistent with the specific heat \cite{Mario}, thermal conductivity\cite{He}, and $\mu$SR\cite{Saghir} measurements in Sn$_{1-x}$In$_{x}$Te, and the scanning tunneling spectroscopy in (Pb$_{0.5}$Sn$_{0.5}$)$_{0.7}$In$_{0.3}$Te (Ref.\cite{Du}).

\begin{figure}[H]
\begin{center}
\includegraphics[width=8cm]{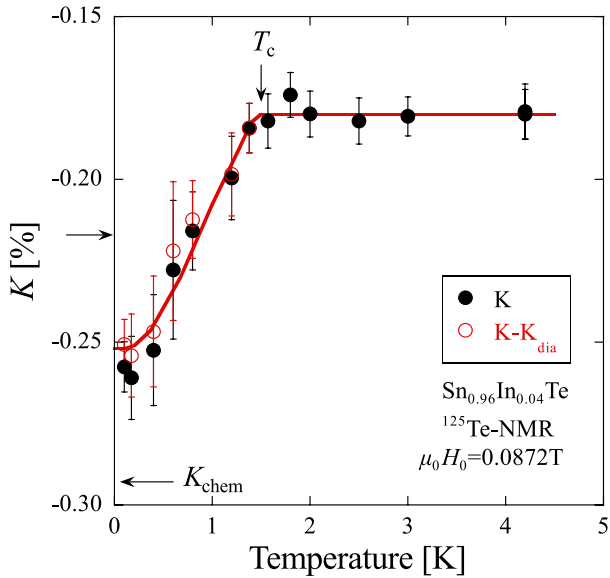}
\caption{Temperature dependence of  the Knight shift $K$ and the corrected value  $K - K_{\rm dia}$ for Sn$_{0.96}$In$_{0.04}$Te under $\mu_0H_0$ =0.0872 T.
The upper arrow indicates the position for the case where one-third of the $K_{\rm s}$ is reduced,
and the lower arrow indicates the position of $K_{chem}$ ($K_{\rm s} = 0$). The curve is a guide for eye.}
\label{K}
\end{center}
\end{figure}

Figure \ref{K}  shows the temperature dependence of the Knight shift $K$, 
which is $T$-independent above $T_{\rm c}$ but decreases below  $T_{\rm c}$.
In the vortex state, one needs to consider a diamagnetic shift $K_{\rm dia}$ ($< 0$) arising from an inhomogeneous field distribution due to the formation of vortex lattices. Namely,  the magnetic field is  position-dependent within the sample,  which can be smaller than  the applied field in some positions.
The position-dependent field $h(\bm{r})$ is calculated using the London model\cite{Zheng},
\begin{equation}
h(\bm{r}) = H \sum_{l, m} \frac{\exp(-G^2_{lm} \xi^2/2) \exp(-i\bm{G}_{lm} \cdot \bm{r})}{1+G^2_{lm} \lambda^2},
\end{equation}
\begin{equation}
\bm{G}_{lm} = 2\pi \sqrt{\frac{H\sin(\beta)}{\phi_0}\left\{ m\bm{x}+\frac{l-m\cos(\beta)}{\sin(\beta)}\bm{y}\right\} },
\end{equation}
where $H$ is the applied field, and $\xi$ is the coherence length.
The summation runs over all reciprocal vortex lattices $\bm{G}_{lm}$, where $\bm{x}$ and $\bm{y}$ are the unit vectors of the vortex lattices, and $\beta$ is the angle between two primitive vortex lattice vectors.
We assumed $\beta = 60^\circ $ and that the $\xi$ and the $\lambda$ are isotropic, reflecting the cubic crystal structure.
The density function of the magnetic field is obtained as $f(h) = \int \delta(h-h(\bm{r}))\ d^3\bm{r}$.
$K_{\rm dia}$ was determined using the peak position of the convolution of the $f(h)$ and the spectrum in the normal state approximated by a Gaussian function.
We used $\mu_0H_{\rm c2}$ = 0.43 T which gives  $\xi$= 27.7 nm from the relation $H_{c2} = \phi_0/2\pi \xi^2$ (Ref. \cite{deGennes}).
The open circles in Fig. \ref{K} show the corrected Knight shift $K - K_{\rm dia}$.
In the figure, the position of  $K_{chem}$ =-0.293\% is marked by the arrow, which is the origin for $K_{\rm s}$.   For a spin triplet state with a cubic crystal structure, a reduction $K_{\rm s}$/3 is expected, whose position is marked by the arrow at $K$=-0.217\%.
Clearly, $K - K_{\rm dia}$ at $T$=0 goes far below  this position. In fact, $K_{\rm s}(T=0)/K_{\rm s}(T=T_{\rm c}) = 0.36 \pm 0.10$ is found. Namely, the reduction is about two-thirds of the total spin Knight shift.
This  result indicates that a spin-singlet superconducting state is realized in the polycrystalline sample of Sn$_{0.96}$In$_{0.04}$Te studied here.

%
\section{Discussions}



We make a few comments on the results and the connection to the topological superconductivity seen in Cu$_x$Bi$_2$Se$_3$. First, we note that even if we use the larger $\mu_0H_{\rm c2}(0)$ = 0.53 T from the Werthamer-Helfand-Hohenberg fitting, our conclusion does not change.
In this case, $K_{\rm dia}(0) = -0.008\%$ and $K_{\rm s}(0)/K_{\rm s}(T_{\rm c}) = 0.39 \pm 0.1$.
Second, the finite $K_{\rm s}$   even at $T$ = 0 can be explained by the scattering due to spin-orbit interaction \cite{Appel}, as seen in many BCS superconductors with large spin-orbit coupling such as  Sn and Hg \cite{Douglas}. A finite $K_{\rm s}$ was also found in Cu$_x$Bi$_2$Se$_3$ when the magnetic field was applied along the $d$-vector direction \cite{Matano}. Thirdly, the isotropic superconducting state found here  is consistent with the  quasi-localized impurity bound states due to In-doping  \cite{MaedaSnTe}. 
As the impurity bound state has no translational symmetry,  a wave number-independent gap is natural.

The results obtained in this work do not support the notion that the superconductivity in Sn$_{0.96}$In$_{0.04}$Te is topological. For a material with spatial inversion  and time reversal symmetry, 
a sufficient conditions for topological superconductivity have been established; namely, 
the parity of the wave function for electron pairs in the superconducting state is odd \cite{Fu}, and the Fermi surface encloses an odd number of time-reversal-invariant momenta \cite{Fu}. These two conditions are fulfilled in Cu$_x$Bi$_2$Se$_3$ \cite{Matano,Xia}.
 The identification of spin-singlet superconductivity in this work suggests that the superconducting wave function of Sn0.96In0.04Te has an even parity and hence it is likely to be a conventional, topologically trivial superconductor.

Quite often, surface-sensitive probes 
and bulk-sensitive  probes such as NMR give different conclusions \cite{Sasaki-Cu,STM,Chu,Matano}. Sometimes the results are different even among the surface-sensitive probes, as encountered in the studies of Cu$_x$Bi$_2$Se$_3$  \cite{Sasaki-Cu,STM,Chu}.  The situation is also true for the current compound for which unconventional superconductivity was previously suggested by point-contact spectroscopy \cite{Sasaki}.
We note that, as a consequence of the topological superconductivity in the bulk, a gapless edge state can  appear in the surface which can be seen by surface-sensitive probes.
 However, the presence or absence of a signature for a surface state alone dose not immediately indicate the properties of the bulk. This is because, in addition to the technical issues \cite{Chu},   the surface has additional complications.
Due to the broken inversion-symmetry on the surface and the strong spin-orbit coupling, parity mixing occurs on the surface. Thus, even the bulk of Cu$_x$Bi$_2$Se$_3$ has an odd parity, $s$-wave component that can be seen on the surface \cite{Mizushima}. The opposite situation, as in the case of Sn$_{1-x}$In$_x$Te, is also possible. 
In the present case, there is another possibility that may reconcile the  different results of NMR  and the previous point-contact spectroscopy \cite{Sasaki}. That is, the sample purity is different in the two measurements. The sample used in the previous study is a single crystal and has less disorder \cite{Mario}, while the sample used in NMR has more disorder as evidenced by the extremely low residual-resistivity ratio ($\sim$1.3). It was reported previously that point-contact spectroscopy depends strongly on the degree of disorder of the samples\cite{Mario}. In more disordered crystals, no zero-bias peak was observed \cite{Mario}.

\section{Conclusion}
In summary, we have performed $^{125}$Te-NMR in polycrystalline samples of the doped TCI, Sn$_{1-x}$In$_x$Te.
$K_{\rm chem}$ was determined to be -0.293\% from the $K$ - $(T_1T)^{-1/2}$ plot with various $x$'s, and determined the spin Knight shift $K_{\rm s}$ for the $x$=0.04 sample.
The FWHM of the $^{125}$Te-NMR spectra of Sn$_{0.96}$In$_{0.04}$Te was $T$ independent
 below $0.5 T_{\rm c}$, which indicates a fully-gapped superconducting state.
$K_{\rm s}(T=0)/K_{\rm s}(T=T_{\rm c})$ reached $0.36 \pm  0.10$,
 which is much smaller than the limiting value of 2/3 for a spin-triplet state in a polycrystal sample with a cubic crystal structure. 
These results indicate that the measured polycrystaline sample of Sn$_{0.96}$In$_{0.04}$Te is  a  spin-singlet superconductor.

\begin{acknowledgments}
We thank Zhiwei Wang for the Hall coefficient measurements and S. Katsube, K. Segawa, and S. Kawasaki for help with some of the measurements, and acknowledge partial support by MEXT Grant No. 15H05852 (Topological Materials Science) and JSPS Grants
No. 16H0401618 and No. 17K14340, as well as by NSFC (Grant No.   11634015) and DFG (CRC1238 "Control and Dynamics of Quantum Materials", Project A04).
\end{acknowledgments}

\end{document}